\documentclass[3p,times,twocolumn]{elsarticle}

\usepackage{ecrc}
\usepackage{multirow}
\usepackage{subfigure}
\usepackage{amssymb}
\usepackage{amsmath}
\usepackage{blkarray}
\usepackage{multirow}
\usepackage{mathtools}
\usepackage{relsize}

\volume{00}

\firstpage{1}

\journalname{Nuclear Physics B Proceedings Supplement}

\runauth{}

\jid{nuphbp}

\jnltitlelogo{Nuclear Physics B Proceedings Supplement}

\usepackage[figuresright]{rotating}

\def\lesssim{\mathrel{\hbox{\rlap{\hbox{\lower5pt\hbox{$\sim$}}}\hbox{$<$}}}}
\def\gtrsim{\mathrel{\hbox{\rlap{\hbox{\lower5pt\hbox{$\sim$}}}\hbox{$>$}}}}

\catcode`\@=11
\def\lsim{\mathrel{\mathpalette\@versim<}}
\def\gsim{\mathrel{\mathpalette\@versim>}}
\def\@versim#1#2{\vcenter{\offinterlineskip
\ialign{$\m@th#1\hfil##\hfil$\crcr#2\crcr\sim\crcr } }}
\catcode`\@=12
\def\be{\begin{equation}}   %
\def\ee{\end{equation}}   %
\def\bea{\begin{eqnarray}}   %
\def\eea{\end{eqnarray}}   %
\def\baa{\begin{align}}   %
\def\eaa{\end{align}}   %

\begin{document}

\begin{frontmatter}



\dochead{HIP-2014-22/TH}

\title{Triplet Extended MSSM: Fine Tuning vs Perturbativity \& Experiment}


\author[label2]{Priyotosh Bandyopadhyay}
 \ead{priyotosh.bandyopadhyay@helsinki.fi}

\author[label2]{Stefano Di Chiara\corref{cor1}}
 \ead{stefano.dichiara@helsinki.fi}

 \author[label2]{Katri Huitu}
 \ead{katri.huitu@helsinki.fi}

 \author[label2]{Asl{\i} Sabanc{\i} Ke\c{c}eli}
 \ead{asli.sabanci@helsinki.fi}

\address[label2]{Department of Physics, University of Helsinki and Helsinki Institute of Physics,
P.O.Box 64 (Gustaf H\"allstr\"omin katu 2), FIN-00014, Finland}

\cortext[cor1]{Corresponding author}

\begin{abstract}
In this study we investigate the phenomenological viability of the $Y=0$ Triplet Extended Supersymmetric Standard Model (TESSM) by comparing its predictions with the current Higgs data from ATLAS, CMS, and Tevatron, as well as the measured value of the $B_s\to X_s \gamma$ branching ratio. We scan numerically the parameter space for data points generating the measured particle mass spectrum and also satisfying current direct search constraints on new particles. We require all the couplings to be perturbative up to the scale $\Lambda_{\rm UV}=10^4$ TeV, by running them with newly calculated two loop beta functions, and find that TESSM retains perturbativity as long as $\lambda$, the triplet coupling to the two Higgs doublets, is smaller than 1.34 in absolute value. For $|\lambda|\gtrsim 0.8$ we show that the fine-tuning associated to each viable data point can be greatly reduced as compared to values attainable in MSSM. Finally, we perform a fit by taking into account 58 Higgs physics observables along with $\mathcal{B}r(B_s\to X_s \gamma)$, for which we calculate the NLO prediction within TESSM. We find that, although naturality prefers a large $|\lambda|$, the experimental data disfavors it compared to the small $|\lambda|$ region, because of the low energy observable $\mathcal{B}r(B_s\to X_s \gamma)$. 
\end{abstract}

\begin{keyword}
Higgs, Triplet Higgs, Supersymmetry

\end{keyword}

\end{frontmatter}


\section{Introduction}
Supersymmetric models remain among the best motivated extensions of the SM. In Minimal Supersymmetric Standard Model (MSSM) the desired Higgs mass can be achieved with the help of radiative corrections for a large mixing parameter, $A_t$, which in turn generates a large splitting between the two physical stops \cite{mssmsd}, and/or large stop soft squared masses. It was shown in \cite{mssmft} that MSSM parameter regions allowed by the experimental data require tuning smaller than $1\%$, depending on the definition of fine-tuning. Such a serious fine-tuning can be alleviated by having additional tree-level contributions to the Higgs mass, given that in MSSM the tree-level lightest Higgs is restricted to be lighter than $m_Z$, so that sizable  quantum corrections are no longer required. In order to have additional contributions to the tree-level lightest Higgs mass, one can extend the MSSM field content by adding a triplet \cite{Espinosa:1991wt,Espinosa:1991gr,DiChiara:2008rg,chargedH,Delgado:2012sm,Delgado:2013zfa,tessm1,Bandyopadhyay:2014tha} chiral superfield.

In light of fine-tuning considerations, here we consider the Triplet Extended Supersymmetric Standard Model (TESSM)\cite{Espinosa:1991wt,Espinosa:1991gr}. The model we consider here possesses a $Y=0$ SU(2) triplet chiral superfield along with the MSSM field content, where the extended Higgs sector generates additional tree-level contributions to the light Higgs mass and moreover may enhance the light Higgs decay rate to diphoton \cite{DiChiara:2008rg,Delgado:2012sm,Delgado:2013zfa,tessm1}. 

To assess the viability of TESSM for the current experimental data, we perform a goodness of fit analysis, by using the results from ATLAS, CMS, and Tevatron on Higgs decays to $ZZ,WW,\gamma\gamma,\tau\tau,b\bar{b}$, as well as the measured $B_s\to X_s \gamma$ branching ratio, for a total of 59 observables. 

\section{The Model} \label{modintr}
The field content of TESSM is the same as that of the MSSM with an additional field in the adjoint of SU$(2)_L$, the triplet chiral superfield $\hat T$, with zero hypercharge ($Y=0$), where the scalar component $T$ can be written as
\be
T=\left(\begin{array}{cc}\frac{1}{\sqrt{2}} T^0 & T^+ \\T^- & -\frac{1}{\sqrt{2}}T^0\end{array}\right)\ .
\ee
The renormalizable superpontential of TESSM includes only two extra terms as compared to MSSM, given that the cubic triplet term is zero:
\bea
W_{\rm TESSM}&=&\mu_T {\rm Tr}(\hat T \hat T) +\mu_D \hat H_d\!\cdot\! \hat H_u + \lambda \hat H_d\!\cdot\! \hat T \hat H_u +\nonumber\\
&&y_t \hat U \hat H_u\!\cdot\! \hat Q - y_b \hat D \hat H_d\!\cdot\! \hat Q- y_\tau \hat E \hat H_d\!\cdot\! \hat L ,
\label{SP}
\eea
where "$\cdot$" represents a contraction with the Levi-Civita symbol $\epsilon_{ij}$, with $\epsilon_{12}=-1$, and a hatted letter denotes the corresponding superfield. The soft terms corresponding to the superpotential above and the additional soft masses can be written similarly\footnote{We use the common notation using a tilde to denote the scalar components of superfields having a SM fermion component.} as
\bea\label{softV}
V_S&=&\left[\mu_T B_T {\rm Tr}(T T) +\mu_D B_D H_d\!\cdot\! H_u + \lambda A_T H_d\!\cdot\! T H_u\right. \nonumber\\
&+& \left.y_t A_t \tilde{t}^*_R H_u\!\cdot\! \tilde{Q}_L + h.c.\right]+m_T^2 {\rm Tr}(T^\dagger T)   \nonumber\\
       &+& m_{H_u}^2 \left|H_u\right|^2  + m_{H_d}^2 \left|H_d\right|^2 + \ldots  \ .
\eea
In the following we assume all the coefficients in the Higgs sector to be real, as to conserve CP symmetry. We moreover choose real vevs for the scalar neutral components, so as to break correctly EW symmetry SU$(2)_L\times$ U$(1)_Y$:
\be\label{vev}
\langle T^0\rangle=\frac{v_T}{\sqrt{2}}\ ,\quad\langle H_u^0\rangle=\frac{v_u}{\sqrt{2}}\ ,\quad\langle H_d^0\rangle=\frac{v_d}{\sqrt{2}}\ ,
\ee
which generate a tree-level contribution to the EW $T$ parameter \cite{Peskin:1991sw,Burgess:1993vc}:
\be
\alpha_e T=\frac{\delta m_W^2}{m_W^2}=\frac{4 v_T^2}{v^2}\ ,\quad v^2=v_u^2+v_d^2\
\ee
with $\alpha_e$ being the fine structure constant. The measured value of the Fermi coupling $G_F$ and the upper bound on the EW parameter $T$ ($\alpha_e T\leq 0.2$ at 95\% CL) \cite{Beringer:1900zz} then impose
\be
v_w^2=v^2+4 v_T^2=\left(\rm 246~GeV\right)^2\ ,\quad v_T \lesssim 5~{\rm GeV}\ .
\ee
Throughout this paper we simply take a small but non-zero fixed value for $v_T$:
\be
v_T= 3\sqrt{2} ~{\rm GeV}\ .
\ee

\section{Higgs Mass \& Direct Search Constraints} \label{HmConst}

After EW symmetry breaking, the stability conditions for the full potential are defined by
\be
\partial_{a_i} V|_{\rm vev} =0\ ,\  \langle a_i\rangle = v_i \ , \  i=u,d,T\ ,
\ee
The conditions above allow one to determine $m^2_{H_u},m^2_{H_d},m^2_{T}$ of the Lagrangian free parameters.
In the limit of large $B_D^2$, which favours EW symmetry breaking \cite{Bandyopadhyay:2014tha}, one can derive an important bound on the mass of the lightest neutral Higgs \cite{Espinosa:1991wt,Espinosa:1991gr}
\be\label{mhbnd}
m^2_{h^0_1}\leq m_Z^2 \left( \cos{2\beta} + \frac{\lambda^2}{g_Y^2+g_L^2} \sin{2\beta} \right)\ ,\ \tan\beta=\frac{v_u}{v_d}\ .
\ee
The result in Eq.~\eqref{mhbnd} shows the main advantage and motivation of TESSM over MSSM: for $\tan\beta$ close to one and a large $\lambda$ coupling it is in principle possible in TESSM to generate the experimentally measured light Higgs mass already at tree-level \cite{DiChiara:2008rg}, which would imply no or negligible Fine-Tuning (FT) of the model. 

\subsection{One Loop Potential}

The one loop contribution to the scalar masses is obtained from the Coleman-Weinberg potential \cite{Coleman:1973jx}, given by
\begin{align}\label{VCW}
V_{\rm CW}=\frac{1}{64\pi^2}{\rm STr}\left[ \mathcal{M}^4
\left(\log\frac{\mathcal{M}^2}{\mu_r^2}-\frac{3}{2}\right)\right],
\end{align}
where $\mathcal{M}^2$ are field-dependent mass matrices in which the fields are not replaced with their vevs nor the soft masses with their expressions at the EW vacuum, $\mu_r$ is the renormalization scale, and the supertrace includes a factor of $(-1)^{2J}(2J+1)$, with the spin degrees of freedom appropriately summed over. The corresponding one loop contribution to the neutral scalar mass matrix, $\Delta{\cal M}^2_{h^{0}}$, is given by \cite{DiChiara:2008rg}
\be
(\Delta\mathcal{M}^2_{h^0})_{ij}
=\left.\frac{\partial^2 V_{\rm{CW}}(a)}{\partial a_i\partial a_j}\right|_{\rm{vev}}
-\frac{\delta_{ij}}{\langle a_i\rangle}\left.\frac{\partial V_{\rm{CW}}(a)}{\partial a_i}\right|_{\rm{vev}}, i,j=u,d,T\ ;
\label{1Lmh}
\ee
where the second term in Eq.~\eqref{1Lmh} takes into account the shift in the minimization conditions, and $a_i$ represent the real components of the scalar fields. 

To evaluate the phenomenological viability of TESSM we proceed by scanning randomly the parameter space for points that give the correct light Higgs mass while satisfying the constraints from direct searches of non-SM particles. The region of parameter space that we scan is defined by:
\begin{align}\label{pscan}
&\!\!\!1\leq t_{\beta }\leq 10\ ,\ 5 \,\text{GeV}\leq \left|\mu _D,\mu _T\right|\leq 2 \,\text{TeV} ,\nonumber\\ 
&50 \,\text{GeV}\leq \left|M_1,M_2\right|\leq 1  \,\text{TeV}\ , \left| A_t,A_T,B_D,B_T\right|\leq 2 \,\text{TeV},\nonumber\\ 
&500 \,\text{GeV}\leq m_Q,m_{\tilde{t}},m_{\tilde{b}}\leq 2 \,\text{TeV}\ ,
\end{align}
with the last three parameters being, respectively, the left- and right-handed squark squared soft masses. The value of $\lambda$ at each random point in the parameter space is determined by matching the lightest Higgs mass at one loop to 125.5~GeV. Having implemented the setup outlined above, we scan randomly the parameter space defined in Eq.~\eqref{pscan} and collect 13347 points that satisfy the constraints
\begin{align}
&m_{h_1^0}=125.5\pm 0.1\, {\rm GeV}\ ;\ m_{A_{1,2}},\ m_{\chi^0_{1,2,3,4,5}}\geq  65\,{\rm GeV}\ ;\nonumber\\
&m_{h^0_{1,2}} , m_{h^\pm_{1,2,3}}, m_{\chi^\pm_{1,2,3}}\geq 100\,{\rm GeV} \ ;\ m_{\tilde{t}_{1,2}},m_{\tilde{b}_{1,2}}\geq  650\,{\rm GeV}\ .
\end{align}
\section{Perturbativity vs Fine-Tuning} 
\label{finetuningTESSM}
We  use the two loop beta functions for the dimensionless couplings of the superpotential and the gauge couplings \cite{Bandyopadhyay:2014tha}, and run each coupling from the renormalization scale $\mu_r=m_Z$ to the GUT scale, $\Lambda_{\rm GUT}=2\times 10^{16}$ GeV. Among the 13347 viable points collected with the random scan described in the previous section, only 7332, or about half, retain perturbativity at the GUT scale. Among these points, the maximum value of $\left|\lambda\right|$ is 0.85 (0.84 at one loop). 

\begin{figure}[htb]
\centering
\includegraphics[width=0.46\textwidth]{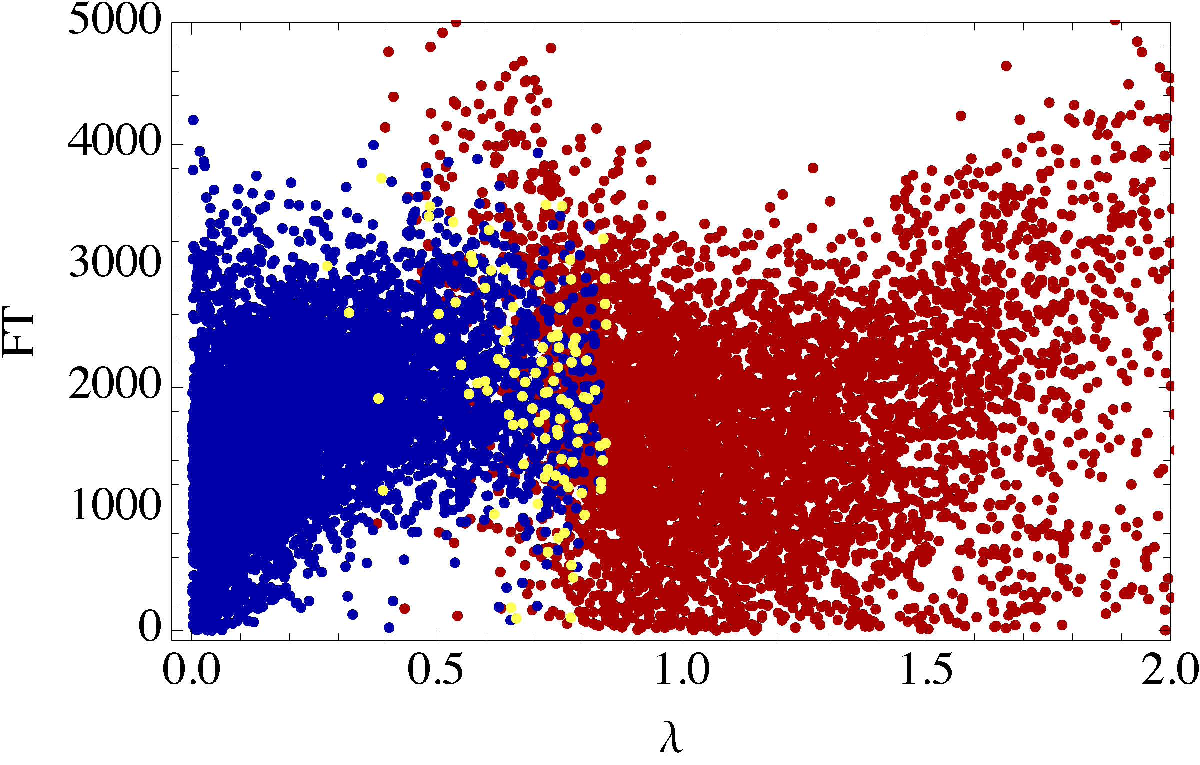}\hspace{0.1cm}
\caption{FT as a function of the triplet coupling $\lambda$: in (red) blue are the (non-perturbative) perturbative points, for which (some) no coupling exceeds $2\pi$ at $\Lambda_{\rm GUT}=2\times 10^{16}$ GeV. In yellow are the points which are perturbative for the two loop but not for the one loop beta functions.}
\label{lambdaFT1}
\end{figure}

A simple estimate of FT in supersymmetry (SUSY) is given by the logarithmic derivative of the EW vev $v_w$ with respect to the logarithm of a given model parameter $\mu_p$ \cite{Ellis:1986yg,Barbieri:1987fn}: this represents the change of $v_w$ for a 100\% change in the given parameter, as defined below:
\begin{align}
&\text{FT}\equiv \frac{\partial  \log  v_w^2}{\partial  \log  \mu_p ^2 \left(\Lambda\right)  }\ ,\ \beta _{\mu_p ^2}=16 \pi ^2 \frac{d\mu_p ^2}{d\text{logQ}}\ ,\nonumber\\
&\mu_p ^2 \left(\Lambda\right) =\mu_p ^2 \left(M_Z\right)+\frac{\beta _{\mu_p ^2}  }{16 \pi ^2} \log   \left(\frac{\Lambda}{M_Z}\right)\ ,
\end{align}
where in parenthesis is the renormalisation scale of $\mu_p$. In MSSM $v_w$ shows its strongest dependence on $m_{H_u}^2$, which therefore produces also the largest value of FT: this is understandable given that the physical light Higgs is mostly of up type. The value of FT in $m_{H_u}^2$, which we calculate by deriving the one loop beta function of $m_{H_u}^2$, indeed happens to be largest in TESSM as well:
\begin{align}\label{FTdef}
&{\rm FT} =\frac{\log\left(\Lambda /M_Z\right)}{16\pi \partial_{v_w^2}m_{H_u}^2}\left[6 y_t^2 A_t^2+3 \lambda^2 A_T^2+3 \lambda ^2 m_{H_d}^2+3 \lambda ^2 m_T^2\right.\nonumber\\
&+3 \lambda ^2 m_{H_u}^2-2 g_Y^2  M_1^2-6 g_L^2 M_2^2 + 6 m_Q^2 y_t^2 + .6 m_{\tilde{t}}^2 y_t^2+6 m_{H_u}^2 y_t^2\nonumber\\
&+\left. g_Y^2 \left(3 m_{\tilde{b}}^2-m_{H_d}^2-3 m_L^2+3 m_Q^2-6 m_{\tilde{t}}^2+m_{H_u}^2+3 m_{\tilde{\tau}}^2\right)\right] .
\end{align}
In Fig.~\ref{lambdaFT1} we present the value of FT evaluated at $\Lambda_{\rm GUT}$, where in blue are the perturbative points, in yellow are 102 points that are non-perturbative only at one loop, while in red are the nonperturbative points: it is clear that while values of $\lambda(M_Z)\sim 1$ indeed produce smaller FT, these large values also drive TESSM into a non-perturbative regime. 

\begin{figure}[htb]
\centering
\includegraphics[width=0.38\textwidth]{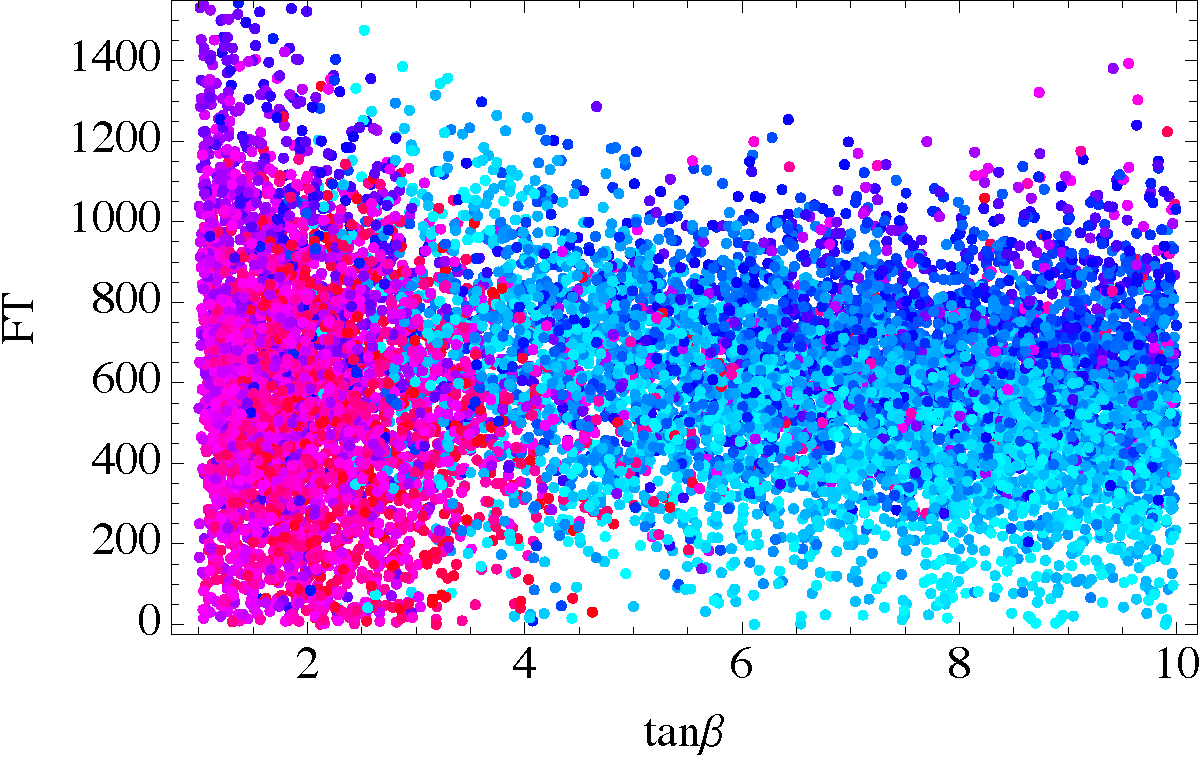}\hspace{0.1cm}
\includegraphics[width=0.08\textwidth]{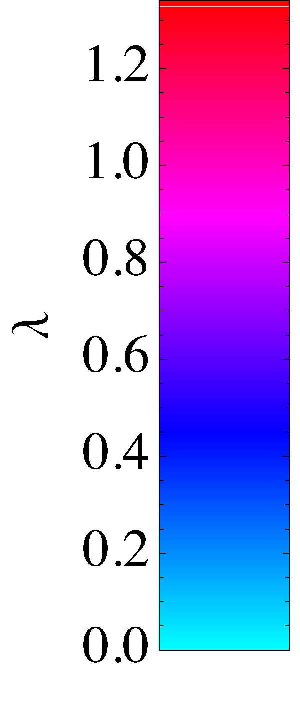}\hspace{0.1cm}
\caption{FT as a function of $\tan\beta$: the region of small $\tan\beta$ and small FT is accessible only for values of $\lambda>0.8$.}
\label{lambdaFT2}
\end{figure}

Taking a cutoff scale as high as the GUT scale, though, is less justifiable for TESSM than for MSSM, given that the triplet in the particle content spoils the unification of the gauge couplings at $\Lambda_{\rm GUT}$. In the following analysis we choose a less restrictive cutoff scale, $\Lambda_{\rm UV}=10^4$ TeV, which is approximately the highest scale tested experimentally through flavor observables \cite{Beringer:1900zz}. Among the 13347 scanned viable data points, 11244 retain perturbativity at $\Lambda_{\rm UV}$, featuring $|\lambda|\leq 1.34$. In Fig.~\ref{lambdaFT2} we plot the FT associated to each of these viable points in function of $\tan\beta$, with a colour code showing the corresponding value of $|\lambda|$. Values of $\tan\beta$ close to 1 can be reached only for large values of $|\lambda|$ (greater than about 0.8) where the corresponding FT can be considerably smaller than for small values of $|\lambda|$, naively associated to MSSM-like phenomenology. 

\begin{figure}[htb]
\includegraphics[width=0.46\textwidth]{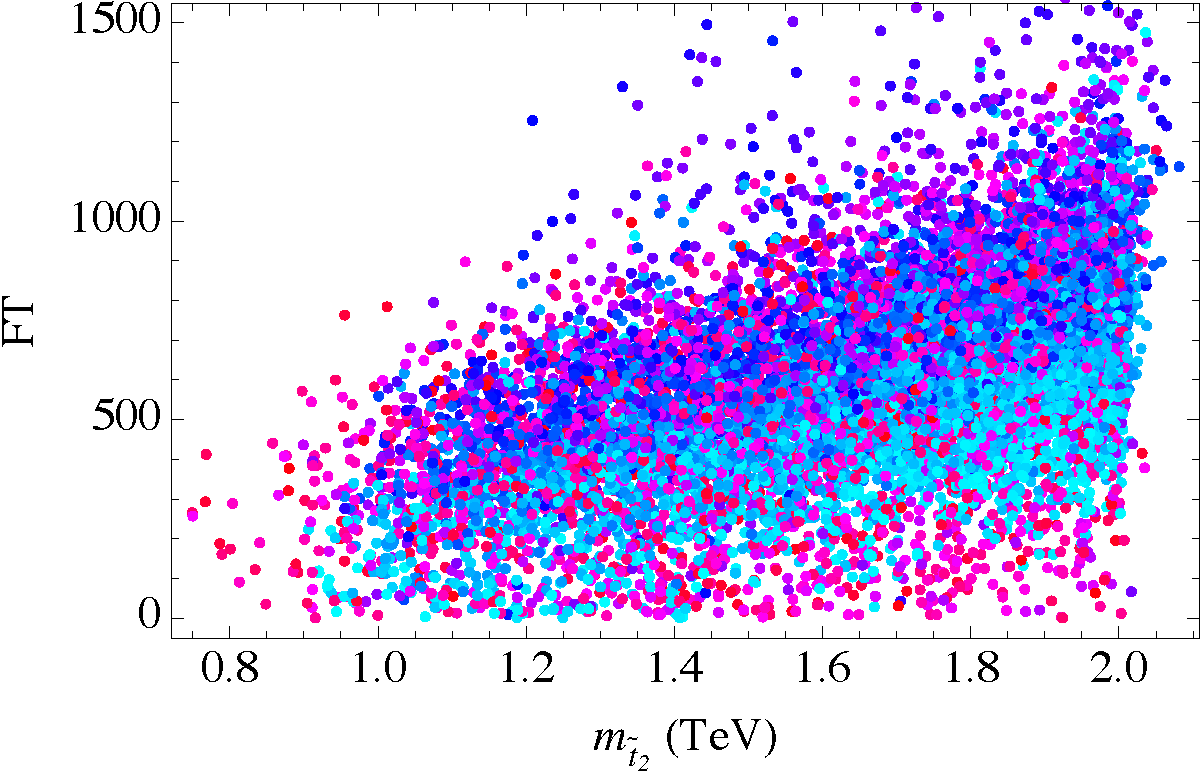}\hspace{0.1cm}
\includegraphics[width=0.46\textwidth]{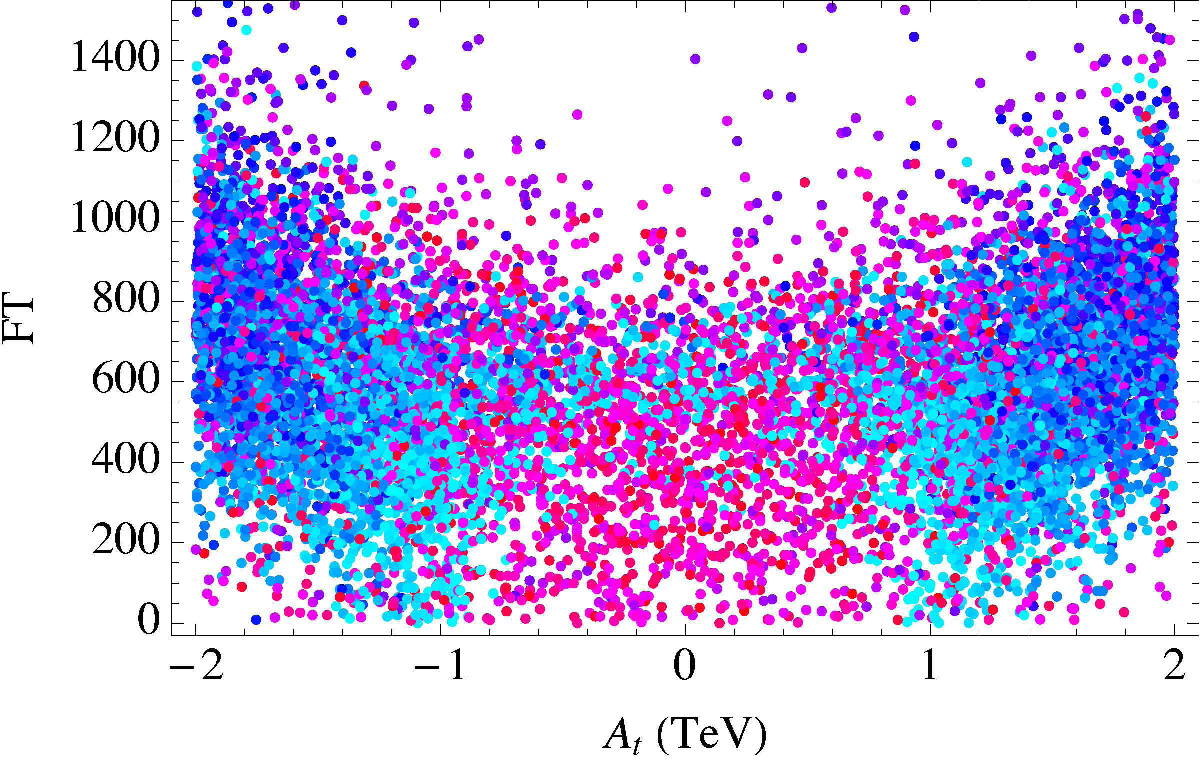}\hspace{0.1cm}
\caption{FT as a function, respectively, of the heavier stop mass $m_{\tilde{t}_2}$ (top panel) and the cubic stop coupling $A_t$ (bottom panel).}
\label{lambdaFT3}
\end{figure}

In Fig.~\ref{lambdaFT3}, FT is plotted both as a function of the heavier stop mass and of $A_t$: the viable region of small $\left|A_t\right|$ and small FT, like that of small $\tan\beta$, is accessible only for large values of $|\lambda|$, greater than about 0.8, where $m_{\tilde{t}_2}$ could be large.

\section{Higgs Physics at LHC}\label{Hphy}
\begin{figure}[htb]
\includegraphics[width=0.46\textwidth]{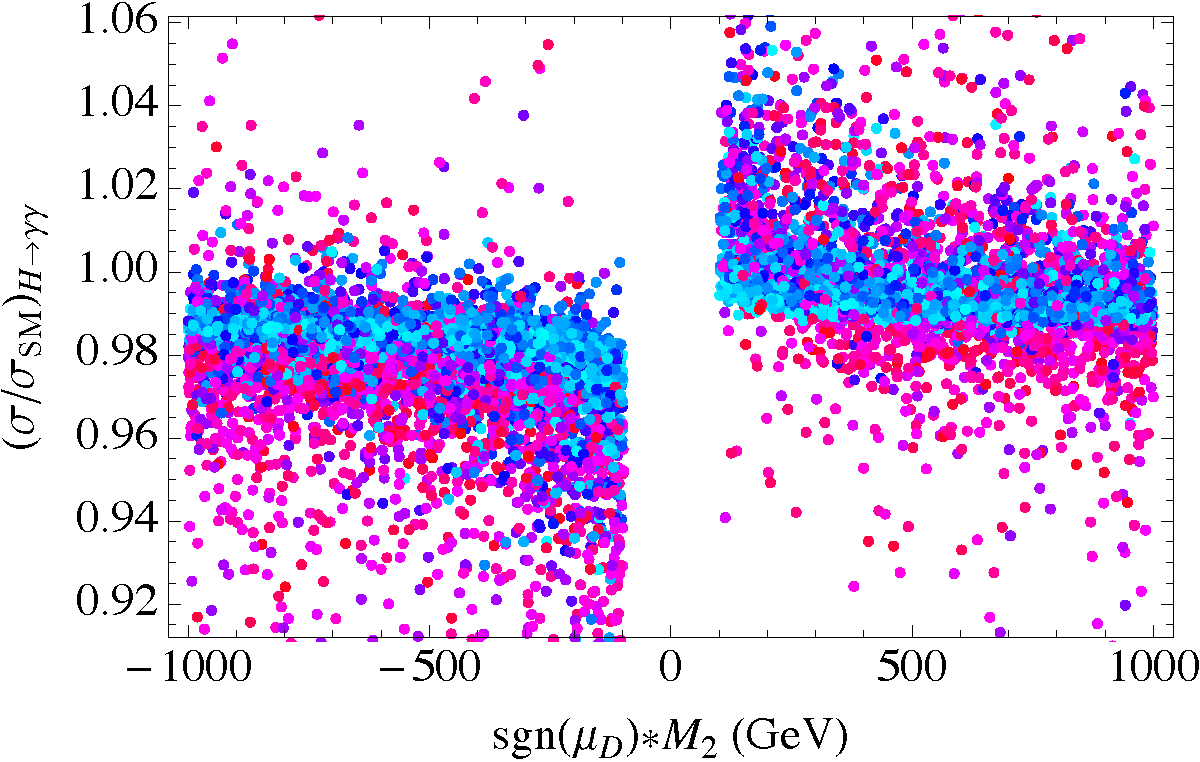}\hspace{0.1cm}
\caption{Higgs decay rate to diphoton of the TESSM relative to the SM as a function, respectively, of ${\rm sign} (\mu_D)\times M_2$.}
\label{H2gamma1}
\end{figure}

The light Higgs linear coupling terms that mimic the TESSM contributions to Higgs physics at LHC can be written as
\begin{align}
&{\cal{L}}_{\textrm{eff}} = a_W\frac{2m_W^2}{v_w}hW^+_\mu W^{-\mu}+a_Z\frac{m_Z^2}{v_w}hZ_\mu Z^\mu \\
&-\sum_{\psi=t,b,\tau}a_\psi\frac{m_\psi}{v_w}h\bar{\psi}\psi-a_\Sigma\frac{2m_\Sigma^2}{v_w}h\Sigma^* \Sigma-a_S\frac{2m_S^2}{v_w}hS^+ S^-.\nonumber
\label{efflagr}
\end{align}
The production cross sections and decay rates for tree-level processes in TESSM are straightforwardly derived by rescaling the corresponding SM result with the squared coupling coefficient of the final particles being produced. For loop induced processes the calculation is more involved. By using the formulas given in \cite{Gunion:1989we} we can write
\be\label{hgamgam}
\Gamma_{h\rightarrow \gamma\gamma}= \frac{\alpha_e^2 m_{h}^3}{256 \pi^3 v_w^2}\left| \sum_i  N_i e^2_i a_i F_{i} \right|^2,
\ee
where the index $i$ is summed over the SM charged particles plus $S^\pm$, $N_i$ is the number of colours, $e_i$ the electric charge in units of the electron charge, and the factors $F_{i}$ are defined in \cite{Gunion:1989we}. We account for the contribution to Higgs decays to diphoton of the charged non-SM particles in TESSM by defining
\be\label{aSd}
a_S \equiv -3 \left[ \sum^3_i \left(F_{h_i^\pm}+F_{\chi_i^\pm}\right)+\sum^2_j \left(\frac{4}{3} F_{\tilde{t}_j}+\frac{1}{3} F_{\tilde{b}_j}\right)\right]\ .
\ee
Similarly to $a_S$ for the two photon decay, $a_\Sigma$ accounts for the contribution on non-SM particles to the light Higgs decay rate to two gluons, and is defined by
\be\label{aSigmad}
a_\Sigma \equiv -3 \sum^2_{j=1} \left(F_{\tilde{t}_j}+F_{\tilde{b}_j}\right)\ .
\ee
We furthermore impose the most stringent limit on the mass of a heavy SM-like Higgs,  $m_{h^0}>770$ GeV,  from the gluon-gluon fusion Higgs production, subsequently decaying to $ZZ$ \cite{CMS:2013ada}. We find this experimental constraint to hold for 10957 out of the 11244 viable data points that already satisfy perturbativity constraints. In Fig.~\ref{H2gamma1} we show the value of the Higgs decay rate to diphoton for TESSM  relative to the SM one, as a function of $ \text{sign}\left(\mu_D\right)\times M_2$, the soft wino mass parameter times the sign of the superpotential doublet mass parameter. The colour code, given in Fig.~\ref{lambdaFT2}, shows the $\left|\lambda\right|$ value corresponding to the plotted data point.  

\section{$\mathcal{B}r(B_s\to X_s \gamma)$ in TESSM}\label{btsgsec}

It has been pointed out in Ref. \cite{Btomumu} that the branching ratio of the flavour changing decay $B_s\rightarrow X_s \gamma$ plays a very important role in constraining the viable parameter space of MSSM especially for low $\tan\beta$. For the numerical analysis we calculate \cite{LOSMCH,NLOSUSYbsg,wilsonrun}, at the next to leading order (NLO) and within TESSM, the values of $\mathcal{B}r(B_s\to X_s\gamma)$ corresponding to each of the scanned 10957 viable data points.

\begin{figure}
\centering
\includegraphics[width=0.46\textwidth]{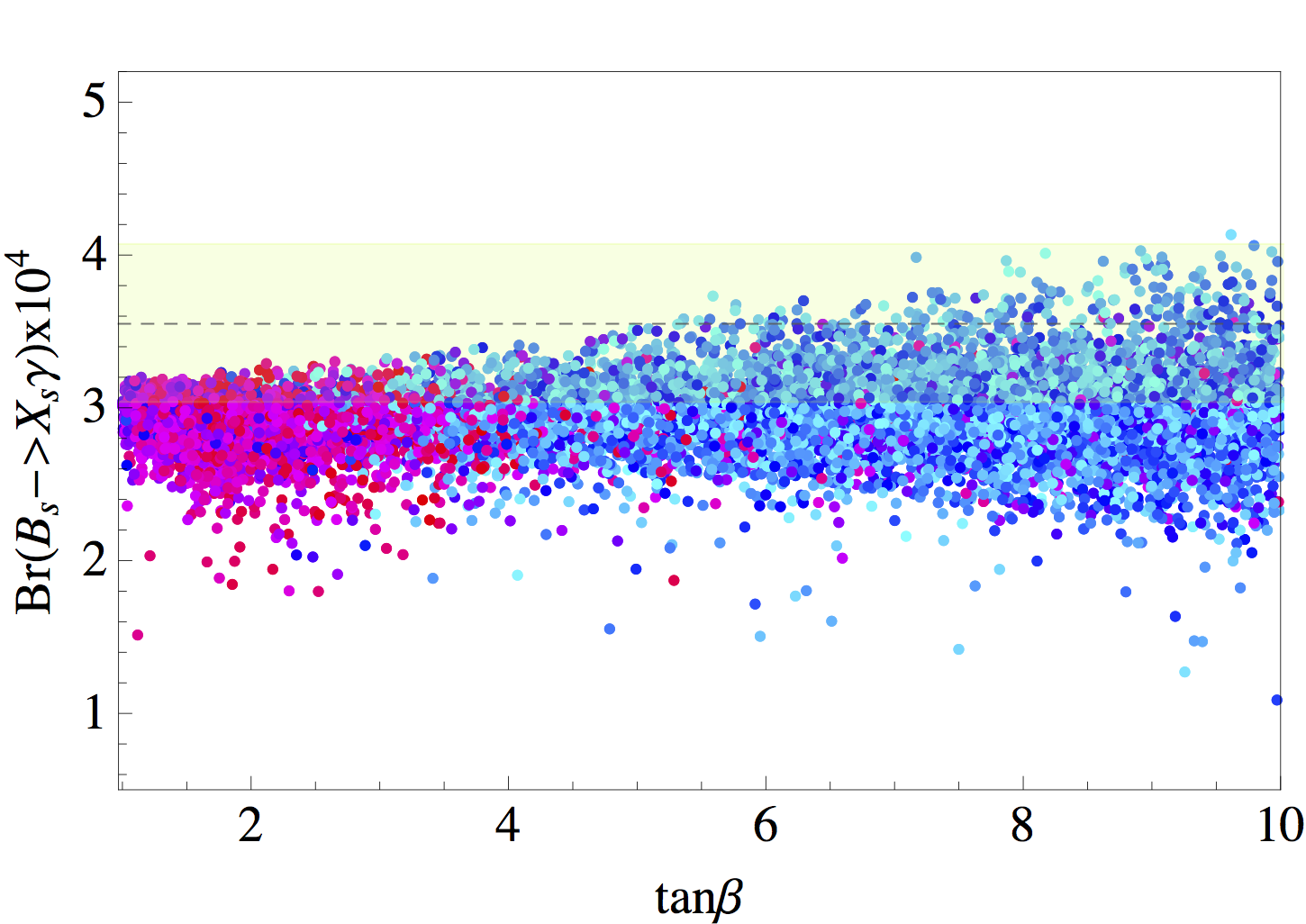}
\caption{The values of $\mathcal{B}r(B_s\to X_s\gamma$) for the allowed data points as a function of $\tan\beta$. The yellow band represents the viable region at $2\sigma$ CL around the experimental value of $\mathcal{B}r(B_s\to X_s\gamma)$.}\label{bsgtan}
\end{figure}

We illustrate the $\tan \beta$ dependence of $\mathcal{B}r(B_s\to X_s\gamma)$, plotted in Fig.~\ref{bsgtan}. For values of $\tan\beta$ close to 10, corresponding to small values of $\lambda$, about half of the data points feature a $\mathcal{B}r(B_s\to X_s\gamma)$ prediction within $\pm2\sigma$ of the experimental value. 
For low $\tan\beta$ values, corresponding to large $\lambda$, the $\mathcal{B}r(B_s\to X_s\gamma)$ values associated to the viable data points sit mostly below the lower $2\sigma$ bound, and for no point the prediction actually matches the experimental value. 

\section{Goodness of Fit to LHC Data \& Conclusions}
\label{fitsect}

To determine the experimentally favored values of the free parameters $a_W,a_Z,a_u,a_d,a_S,a_\Sigma$, we minimize the quantity
\be
\chi^2=\sum_i\left(\frac{{\cal O}_i^{\textrm{exp}}-{\cal O}_i^{\textrm{th}}}{\sigma_i^{\textrm{exp}}}\right)^2,
\label{chi2}\ee
where $\sigma_i^{\textrm{exp}}$ represent the experimental uncertainty, while the observables ${\cal O}_i^{\textrm{exp}}$ correspond to the signal strengths for Higgs decays to $ZZ$, $W^+W^-$, $\tau^+\tau^-$, $b\bar{b}$, as well as all the topologies of decays to $\gamma\gamma$, respectively measured by ATLAS \cite{ATLAS:2013nma,ATLAS:2013wla,ATLAS:2012dsy,ATLAS:2012aha,ATLAS:2013oma} and CMS \cite{CMS:xwa,CMS:bxa,CMS:utj,CMS:2014ega}, and by Tevatron for decays to $W^+W^-$ and $b\bar{b}$ \cite{Aaltonen:2013kxa}. 

\begin{figure}[htb]
\includegraphics[width=0.46\textwidth]{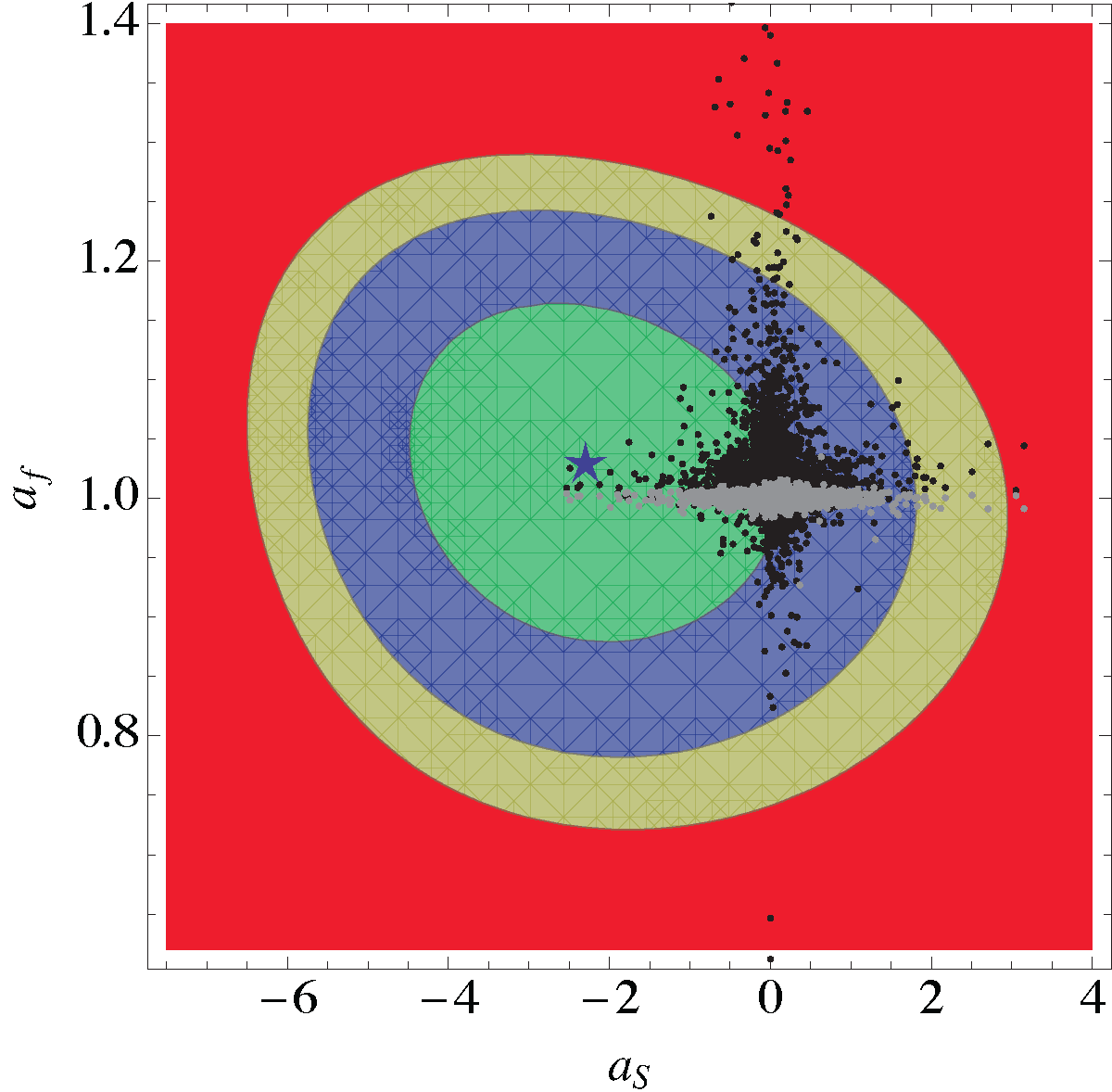}\hspace{0.1cm}
\caption{Viable regions at the 68\%, 95\%, 99\% CL in the coupling coefficients $a_S,a_f$ plane passing through the optimal point (blue star), together with the values of $a_u$ (grey) and $a_d$ (black) associated with each viable point.}
\label{aSaf}
\end{figure}

In calculating $\chi^2$ for the TESSM viable data points we include also the $\mathcal{B}r(B_s\to X_s \gamma)$ observable. Assuming a total of four free parameters ($a_f,a_S,a_\Sigma$, plus one more to fit $\mathcal{B}r(B_s\to X_s \gamma)$), the viable data point featuring minimum $\chi^2$ has
\be
\chi^2_{min}/d.o.f.=1.01\, ,\ d.o.f.=55\,,\ p\left(\chi^2>\chi^2_{min}\right)=46\%\ .
\ee 
This result should be compared with the SM one for the same set of observables:
\be
\chi^2_{min}/d.o.f.=0.99\, ,\ d.o.f.=59\,,\ p\left(\chi^2>\chi^2_{min}\right)=50\%\, .
\ee
We notice that the goodness of fit of TESSM is comparable, although smaller, to that of the SM. It is important to realize, however,  that the quoted $p$ values are only indicative of the viability of TESSM and SM relative to one another. In Fig.~\ref{aSaf} we plot the 68\%, 95\%, 99\% CL viable regions (respectively in green, blue, and yellow) on the plane $a_S-a_f$ intersecting the optimal point (blue star). On the same plane we plot also the values of $a_u$ (gray dots) and $a_d$ (black dots) along the $a_f$ dimension. 

\begin{figure}[htb]
\centering
\includegraphics[width=0.46\textwidth]{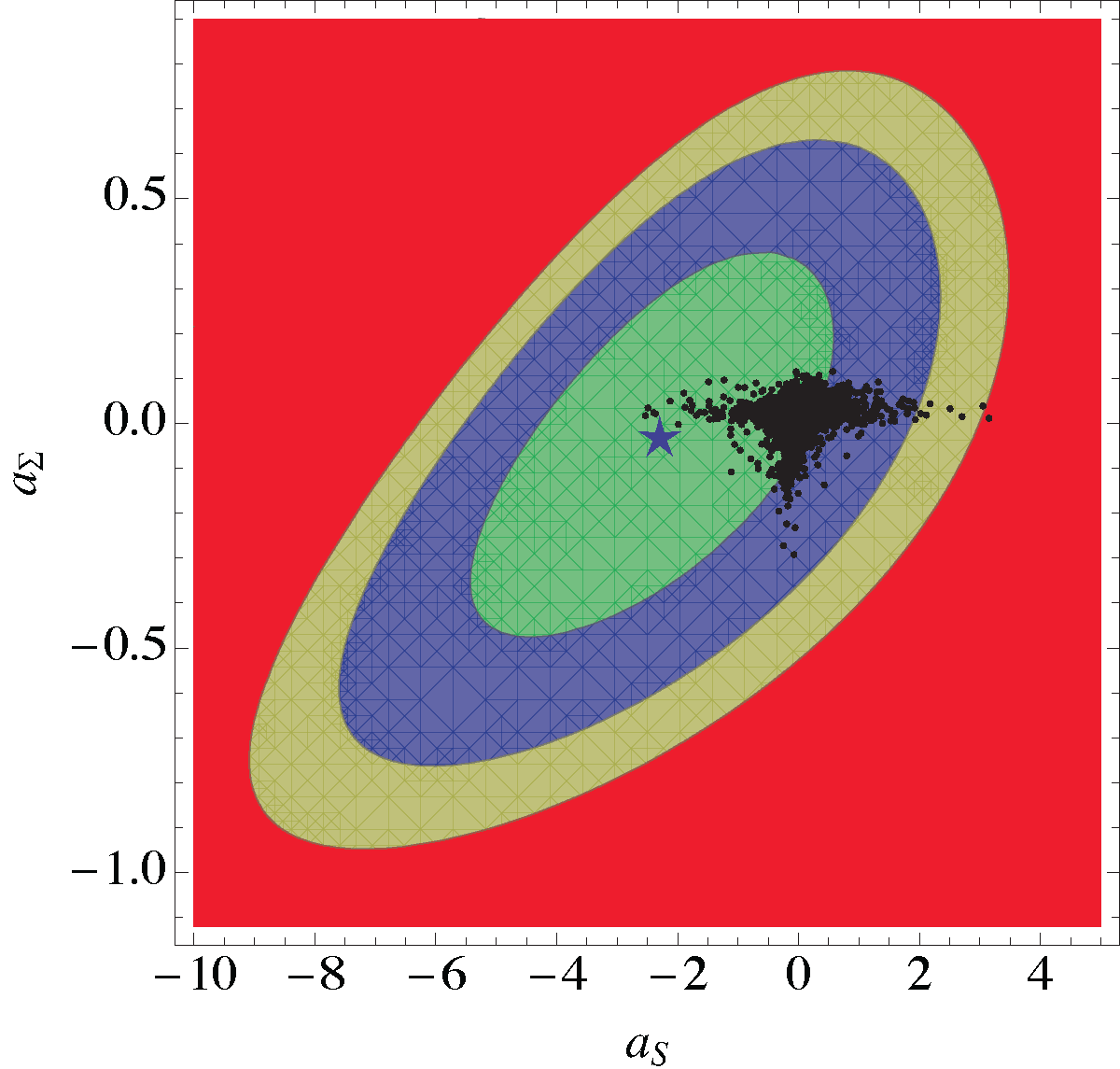}\hspace{0.1cm}
\caption{Viable regions at the 68\%, 95\%, 99\% CL in the coupling coefficients $a_S,a_\Sigma$ plane passing through the optimal point (blue star), together with the corresponding value (black) associated with each viable point.}
\label{aSaSigma}
\end{figure}

In Figs.~\ref{aSaSigma} we plot the 68\%, 95\%, 99\% CL viable regions (respectively in green, blue, and yellow) on the plane $a_S-a_\Sigma$ intersecting the optimal point (blue star), together with the corresponding coupling coefficients values for each viable data point (black).  No viable data point matches the optimal values, as the bulk of data points deviates from it about $1\sigma$ along the $a_S$ axis.

\begin{figure}[htb]
\centering
\includegraphics[width=0.46\textwidth]{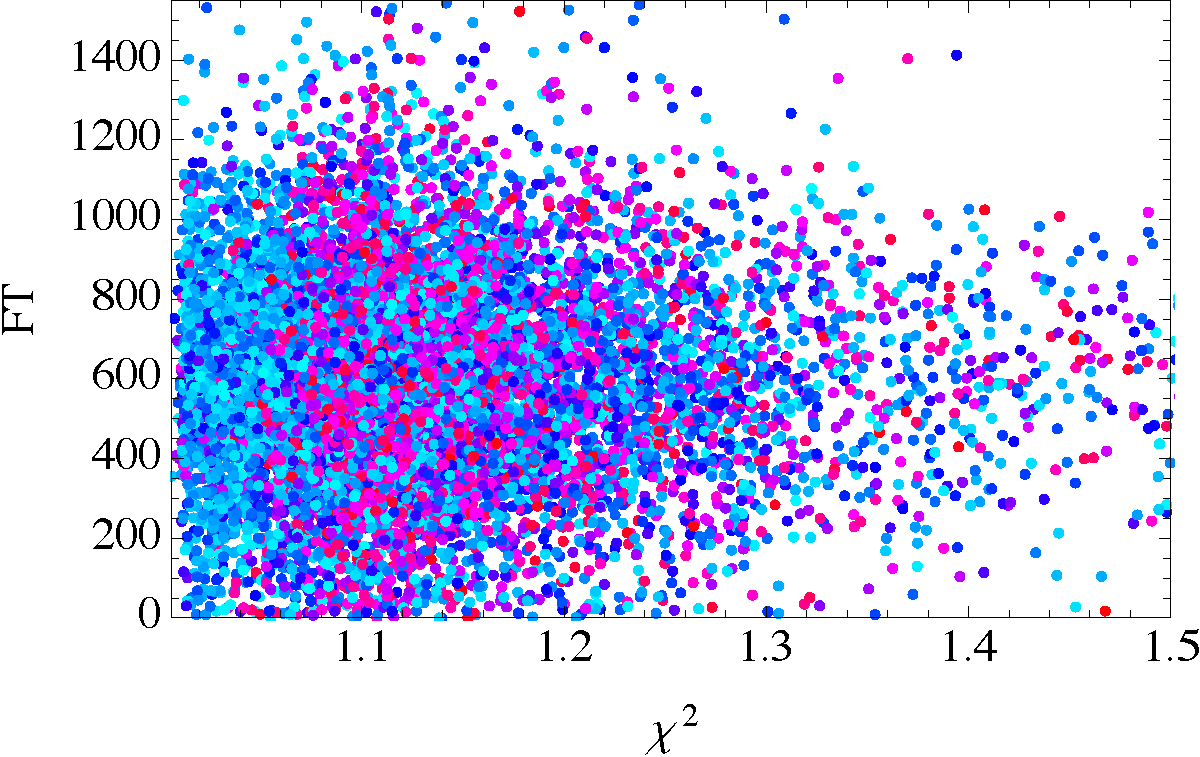}\hspace{0.1cm}
\caption{FT as a function of $\chi^2$ with colour code associated with the absolute value of $\lambda$.}
\label{chi2FT}
\end{figure}

Finally, in Fig.~\ref{chi2FT} we plot the FT for each data point, with the colour code of the absolute value of $\lambda$ defined in Fig.~\ref{lambdaFT2}, as a function of its $\chi^2$ value, which includes the contribution of $\mathcal{B}r(B_s\to X_s \gamma$) defined  in Eq.~\eqref{chi2}. As we can see from Fig.~\ref{bsgtan}, small $|\lambda|$ values more likely satisfy the $\mathcal{B}r(B_s\to X_s \gamma)$ experimental bound. It is important to notice that large absolute values of $\lambda$ are not able to improve the fit to current Higgs physics data enough to compensate for the bad fit to $\mathcal{B}r(B_s\to X_s \gamma$). In a scenario, instead, in which both ATLAS and CMS find a large enhancement with small uncertainty in the next LHC run, the TESSM would achieve a goodness of fit comparable to that of MSSM, with possibly a considerably smaller amount of FT.


\end{document}